\documentclass[10pt]{iopart}
\usepackage[utf8]{inputenc} 

\usepackage{siunitx}
\usepackage{textcomp}
\expandafter\let\csname equation*\endcsname\relax
\expandafter\let\csname endequation*\endcsname\relax
\usepackage{amsfonts,amsmath,amssymb}
\usepackage[dvipsnames]{xcolor}
\usepackage{graphicx}
\usepackage{dcolumn}
\usepackage{bm}
\usepackage{braket}
\usepackage{verbatim}
\usepackage[colorlinks=true,
  urlcolor=blue,
  linkcolor=blue,
  citecolor=blue]{hyperref}
\usepackage{cleveref}

\let\vec\bm
\Crefname{equation}{Eq.}{Eqs.}
\Crefname{figure}{Fig.}{Figs.}
\Crefname{tabular}{Tab.}{Tabs.}
\Crefname{table}{Tab.}{Tabs.}
\Crefname{table}{Tab.}{Tabs.}

\newcommand{\zu}{z}
\newcommand{\zd}{\tilde{z}}
\newcommand{\Zu}{Z}
\newcommand{\Zd}{\tilde{Z}}

\begin{document}
\title{Skyrmion Ground States of Rapidly Rotating Few-Fermion Systems}
\author{L~Palm $^{1,4}$, F~Grusdt $^{2,3}$ and P~M~Preiss $^1$}
\address{$^1$ Physikalisches Institut der Universit\"at Heidelberg, Im Neuenheimer Feld 226, 69120 Heidelberg, Germany\\
$^2$ Department of Physics and Arnold Sommerfeld Center for Theoretical Physics (ASC), Ludwig-Maximilians-Universit\"at M\"unchen, Theresienstr. 37, M\"unchen D-80333, Germany\\  
$^3$ Munich  Center  for  Quantum  Science  and  Technology  (MCQST),  Schellingstraße  4,  80799  M\"unchen,  Germany\\
$^4$ Current Address: James Franck Institute and Department of Physics, University of Chicago, Chicago, IL 60637, USA}
\ead{lpalm@uchicago.edu}

\begin{abstract}
We show that ultracold fermions in an artificial magnetic field open up a new window to the physics of the spinful fractional quantum Hall effect. We numerically study the lowest energy states of strongly interacting few-fermion systems in rapidly rotating optical microtraps.
We find that skyrmion-like ground states with locally ferromagnetic, long-range spin textures emerge. To realize such states experimentally, rotating microtraps with higher-order angular momentum components may be used to prepare fermionic particles in a lowest Landau level. We find parameter regimes in which skyrmion-like ground states should be accessible in current experiments and demonstrate an adiabatic pathway for their preparation in a rapidly rotating harmonic trap. The addition of long range interactions will lead to an even richer interplay between spin textures and fractional quantum Hall physics.
\end{abstract}

\date{\today}

\submitto{\NJP}
\maketitle

\section{Introduction}

The correlated quantum states underlying the fractional quantum Hall (FQH) effect are prime examples of topological phases of matter.
These states cannot be classified in the usual framework of symmetry breaking, but are characterized by patterns of long-range entanglement. 
Such topological states may exhibit intriguing physical properties such as excitations carrying fractional charge or fractional exchange statistics, which cannot be found in ordinary matter.

FQH states were first discovered in two-dimensional electron gases where strong magnetic fields spin-polarize the sample \cite{Laughlin1983,Jain2007}. It was soon realized, however, that for certain semiconductors like GaAs, the Zeeman splitting is small and the electron spin plays an important role in the minimization of interaction energy \cite{Halperin1983}. When multiple spin configurations become degenerate with each other, the samples develop a behavior referred to as quantum Hall ferromagnetism \cite{Yang1994}. In this regime, the spinful electron gas can host stable excitations with unusual spin textures, such as skyrmions \cite{Rezayi1987,Haldane1988,Lee1990,Rezayi1991,Sondhi1993,Fertig1994,Wu1994} carrying a non-zero topological charge. These structures have been studied extensively, but their exact nature and role in strongly correlated systems remain an open question \cite{Young2012,Balram2015}.

A promising experimental approach to this problem is to substitute the electron gas for synthetic quantum systems composed of neutral atoms or photons. This procedure offers the opportunity to study the microscopic properties of topological states via single-particle imaging and in the presence of controlled interactions. Major progress in the quantum simulation of non-interacting quantum Hall systems has recently been made with the creation of the necessary artificial gauge fields on a number of experimental platforms, including microwave and optical photons, and ultracold atoms \cite{Aidelsburger2017}. Signatures of topological states in strongly interacting systems have been far more elusive, with the notable exception of rapidly rotating ultracold bosonic systems \cite{Gemelke2010} and the recent realization of two-photon Laughlin states of light \cite{Clark2019}.

Most efforts to date have focused on the realization of FQH states of spin-polarized bosonic particles \cite{Popp2004,Sorensen2005,Gemelke2010,Roncaglia2011,Zhang2016,He2017,Clark2019} which have been shown to exhibit FQH states such as bosonic Laughlin states. It has been recognized theoretically that multi-component Bose-Einstein condensates develop non-trivial spin textures both for slow \cite{Mueller2004,Kasamatsu2003} and rapid rotation \cite{Ho2002,Reijnders2004}. In contrast to such bosonic equivalents, we consider cases closer to the solid state setting and study spinful FQH states of ultracold fermionic atoms and their connection to QH ferromagnetism.

\section{Quantum Hall Ferromagnetism with Ultracold Atoms}
Ultracold atoms differ in several important regards from their solid state counterparts: Typically, there is no spin bath and the global magnetization and often the total spin are conserved quantities that may be controlled experimentally. Moreover, collisions between ultracold atoms are well-described by $s$-wave interactions: Fermionic atoms scatter only with particles in different hyperfine states, resulting in spin-selective contact interactions. This situation is counter to previously explored bilayer systems \cite{Yoshioka1989, Eisenstein1992, Yang1994}, where the intra-layer interaction is usually much stronger than its inter-layer counterpart. It is therefore an open question whether quantum Hall ferromagnetism and skyrmionic excitations as present in Coulomb systems may be observed with ultracold atoms.  

In this work, we answer this question in the affirmative through a numerical study of the ground states of fermionic neutral atoms in rapidly rotating traps. We find that ground states exhibit ferromagentic behaviour, i.e. they maximize their total spin, while nearby states are spin singlets and exhibit skyrmion-like behaviour: They are locally ferromagnetic, but accommodate reversed spins through long-range spin textures. We identify the emerging skyrmionic states by comparison to many-body trial wavefunctions. 

To access spinful quantum Hall states experimentally, we propose to extend the technique of rapidly rotating microtraps, which so far has been developed mostly for bosonic atoms, to include higher-order perturbations. With this approach, lowest-Landau level physics may also be realized for fermionic atoms. 

Our strategy will enable several new types of experiments: The ferromagnetic ground states, which we identify in the sector of maximal spin, represent a new class of states for synthetic quantum systems. Ferromagnetism in itinerant systems occurs in finely tuned scenarios \cite{Stoner1933} and has not been observed conclusively with ultracold atoms \cite{Jo2009, Pekker2011}. The scenario we consider here allows the detailed study of the formation and stability of a few-body itinerant ferromagnet. 

The main focus of our work, however, lies on the skyrmionic spin textures that emerge as ground states of the singlet spin sector. We identify parameter regimes where skyrmionic states may be  realized in existing experiments and describe adiabatic pathways for their preparation. 

\section{Numerical Model}

We numerically study the low-energy states of few particles confined to two dimensions in rapidly rotating traps, as already considered in \cite{Popp2004,Baur2008,Roncaglia2011}.  
In the limit of rapid rotation, this configuration supports Landau-level-like degenerate single-particle manifolds .
The two principal questions we address here are: (1) How can the single-particle band structure be engineered to load fermionic particles into the lowest Landau level, given their Pauli exclusion; and (2) which many-body states form in the limit of rapid rotation?

\subsection{\label{ssec:single}Single-particle picture}
We start by examining the single-particle spectrum of a spin-less atom in a rotating trap, which is formally equivalent to that of a particle in a strong magnetic field. This problem is well known in the literature \cite{Fetter2009} and we only state the most important features.

The single-particle Hamiltonian can be written in a harmonic oscillator basis as 
\begin{equation}
\label{eq:H0sb}
\mathcal{H}_0(\alpha) = \left(2 a^\dagger a + 1 \right) + \alpha \left(b^\dagger b - a^\dagger a \right),
\end{equation}
with the ladder operators $a^\dagger = - \partial_z + \bar{z}/2$, $b^\dagger = - \partial_{\bar{z}} + z/2$ in terms of the complex coordinate $z = x + iy$ and its complex conjugate $\bar{z}$. The rate of rotation $\Omega$ can be expressed in non-dimensional form via the scaling $\alpha = 1 - \Omega/\omega$. All energy and length scales for particles with atomic mass $M$ are given in units $\hbar \omega$ of the radial trap frequency $\omega$ and corresponding harmonic oscillator length $l = \sqrt{\hbar/M\omega}$, that coincides with the magnetic length in the deconfinement limit. 
In the limit of fast rotation, the more natural quantum numbers $|n, m \rangle $ are the Landau level $n = n_a = \langle a^\dagger a \rangle$ and the states angular momentum $m = n_b - n_a = \langle b^\dagger b \rangle - \langle a^\dagger a \rangle$.
Usually it is sufficient to focus on the $n=0$ subspace, known as the lowest Landau level (LLL), as all QH effect physics takes place in a partially filled LLL or can be mapped to it. This is also the case for bosonic atoms, as even at rest they populate the $|0,0\rangle$ state collectively and remain in the LLL as angular momentum is introduced through rotating trap perturbations \cite{Popp2004,Baur2008}. 
For fermions however, the situation is more complicated. States in the resting trap ($\alpha=1$) are filled up to the Fermi level with a singlet per single-particle level, corresponding to occupations of higher LL $n>0$ in the rapidly rotating limit. This is highly undesirable as it prevents clean access to the quantum Hall physics in the LLL.

\begin{figure}[htb]
	\centering{
	\includegraphics[width=0.95\textwidth]{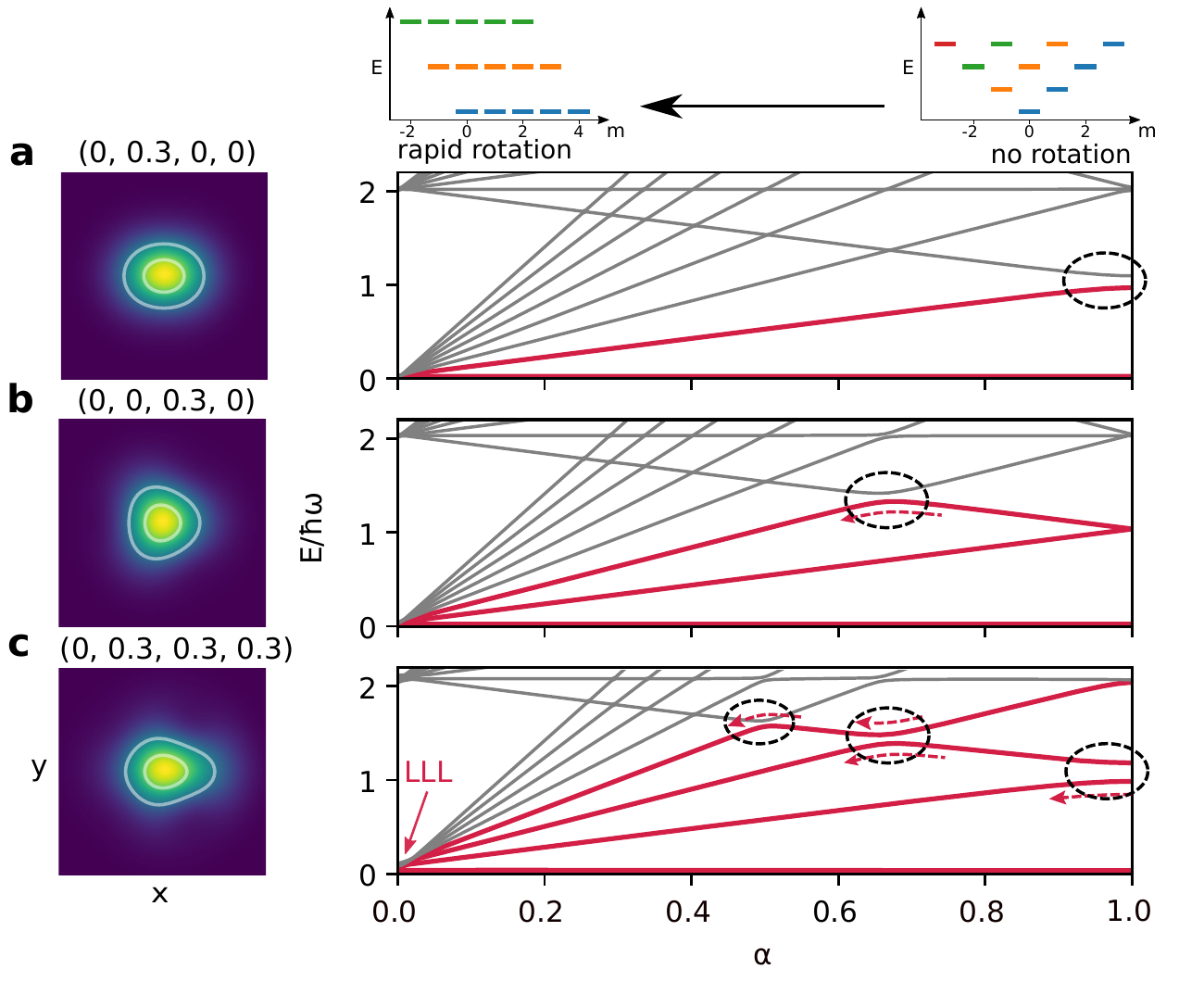}
	\caption{ \label{fig:single-part} Single-particle spectrum. The energy levels of the resting trap (top insert, right) smoothly transform into a set of Landau levels for rapid rotation (top insert, left). Traps with perturbations of higher azimuthal order (real space potentials shown in the left column) can be used to transfer fermions to the lowest Landau level (LLL). With increasingly higher order perturbations, gaps in the spectrum open up (ellipses) and more single-particle paths from the resting system connect to the lowest Landau level (red highlights and arrows). Shown are potentials including perturbations of strength $(\epsilon_1,\epsilon_2,\epsilon_3,\epsilon_4)$ for (a) elliptical, (b) triangular and (c) composite perturbation.}
    }
\end{figure}

We propose to employ higher-order trap perturbations as a natural way to transfer all fermionic particles from a resting trap to the LLL. A rotating optical potential containing azimuthal perturbations up to order $l^*$ can be written as \cite{Baur2008}
\begin{equation}
\label{eq:pert-general}
\mathcal{H}_p(z,t) = \sum_{l=1}^{l^*}  \epsilon_l \left( z^l e^{il\Omega t} + h.c.\right)
\end{equation}
in the lab frame in terms of time $t$, complex coordinate $z$ and $\epsilon_l$ the amplitude of order $l$. The components of this potential couple single particle states with different angular momenta and LL according to the selection rule $\Delta m = m^\prime - m = l$, where the matrix elements $\langle n^\prime,m^\prime |\mathcal{H}_p|n, m\rangle$ are computed from the explicit wavefunctions in real space (see \cite{Roncaglia2011}) through a numerical integration. We work in a rotating frame at frequency $\Omega$ to remove the time-dependence of \Cref{eq:pert-general}. 
By choosing an appropriate superposition of perturbations up to some order $l^*$, gaps open in the spectrum of Hamiltonian $\mathcal{H}_0(\alpha) + \mathcal{H}_p$  wherever levels with angular momentum difference $\Delta m \leq l^*$ cross. This connects $l^*$ levels from the resting trap ($\alpha =1$) to the lowest energy manifold at $\alpha \rightarrow 0$. The desired number of atoms $N_\uparrow = N_\downarrow \leq l^*$ can thus be transferred into the LLL. As shown in \Cref{fig:single-part}, a triangular perturbation ($l=3$) is suited for the transfer of $3+3$ atoms and already at an intermediate rotational frequency of $\alpha = 0.5$ the LLL is reached. We emphasize that after the last single-particle gap has been crossed towards $\alpha=0$, the rotating perturbation can be removed to close the single-particle gaps and a pure elliptic perturbation can be chosen for the final approach to the centrifugal limit. Once all particles are prepared in the LLL, an elliptic perturbation with $l=2$, as considered in the remainder of the paper, is adequate to reach the desired target states, although different orders might be used to reach states of odd total angular momentum.

Using programmable optical elements \cite{Zupancic2016,Papageorge2016}, controlled superpositions of light beams with different orbital angular momentum can be realized, such that adiabatic pathways to the lowest Landau level for up $N_\downarrow=N_\uparrow \approx 10$ atoms using ``designer gaps" in the single-particle spectrum should be feasible. 

\subsection{\label{ssec:mb}Many-body problem}

We now consider an interacting two-component spin mixture in the lowest single-particle eigenstates near the deconfinement limit $\alpha=0$.
The kinetic Hamiltonian \Cref{eq:H0sb} written in the Fock-Darwin basis
\begin{equation}
\mathcal{H}_{0}(\alpha) = \alpha \sum_{m,\sigma} m\, c_{m}^{\sigma\dagger} c_{m}^{\sigma} = \alpha L,
\end{equation}
with $c_{m}^{\sigma\dagger}$ the fermionic creation operator in angular momentum mode $m$ for spin $\sigma$, is diagonal and proportional to the total angular momentum $L$ of the system. 
$S$-wave interactions may be expressed in the orbital basis as
\begin{equation}
\mathcal{H}_{I}(\eta) = \eta \sum_{\{m,\sigma,\sigma^\prime\}} V_{\{m\}}^{\sigma,\sigma^\prime} c_{m_1}^{\sigma~ \dagger} c_{m_2}^{\sigma^\prime~\dagger} c_{m_3}^{\sigma^\prime} c_{m_4}^{\sigma}
\end{equation}
with the dimensionless interaction strength $\eta$ set by the s-wave scattering length $a_s$ and the $z$-confinement harmonic oscillator length $l_z$ as $\eta = \sqrt{8\pi} a_s/l_z$. Angular momentum is conserved through this interaction, which corresponds to the first Haldane pseudo-potential $V_0$ \cite{Haldane1983}.
The coefficients $V_{\{m\}}^{\sigma,\sigma^\prime} = \frac{(m_1+m_2)!\delta_{m_1+m_2,m_3+m_4}}{2^{m_1+m_2}\sqrt{m_1! m_2! m_3! m_4!} } (1-\delta_{\sigma,\sigma^\prime})$ are non-zero only for $\sigma\neq\sigma^\prime$ as only fermions in different hyperfine states interact in $s$-wave scattering.

Our analysis is based on a numerical exact diagonalization of the Hamiltonian $\mathcal{H}= \mathcal{H}_{0}(\alpha) + \mathcal{H}_{I}(\eta) + \mathcal{H}_{p}(\epsilon)$ for small system sizes in the disk geometry. We focus on few-body states with up to four particles per spin state as these are numerically tractable and have several experimental advantages. The critical rate of rotation required to reach the strongly correlated regime increases with particle number, therefore the frequencies for large system sizes are prohibitively close to the deconfinement limit \cite{Fetter2009}. By choosing a small system size, the required rate of rotation is reduced to experimentally more realistic values where small trap anharmonicities and imperfections can be tolerated. A small number of particles is also compatible with  free-space imaging \cite{Bergschneider2018} that can give access to particle-resolved correlation functions.

\section{FQH Ferromagnet and Skyrmions}
We are interested in FQH states in the LLL, which are polynomials in the particles' complex coordinates $z = x + i y$ \cite{Laughlin1983,Jain2007} (up to a Gaussian envelope factor which we omit throughout). The two spin components can be encoded as two different variables $\zu_i$ and $\zd_i$ for spin-up and -down particles. Laughlin's celebrated wavefunction \cite{Laughlin1983} for a spin-polarized system was generalized by Halperin \cite{Halperin1983} to include a spin degree of freedom, which leads to a series of $\Psi_{(m_1,m_2,n)}$ wavefunctions
\begin{equation}
\label{eq:Halperin}
\Psi_{(m_1,m_2,n)} = \prod_{i<j}^{N_\uparrow} (\zu_j - \zu_i)^{m_1} \prod_{l<k}^{N_\downarrow} (\zd_l - \zd_k)^{m_2} \prod_{p,q}^{N_\uparrow,N_\downarrow} (\zu_q - \zd_p)^{n}.
\end{equation}
for different integer exponents $m_1,m_2,n$. 
The filling fraction for a $(m,m,n)$ state is $\nu = \nu^\uparrow + \nu^\downarrow =  2/(m+n)$ and we restrict ourselves to the case of balanced spins $N_\uparrow = N_\downarrow = N$ as well as $m_1 = m_2$ throughout the paper.

Several interesting fermionic states can be deduced from \Cref{eq:Halperin}: A first, trivial state is given by the $\Psi_{(1,1,0)}$ wavefunction, denoting a product of Fermi seas of both components in the lowest LL. A state with vanishing interaction energy is $\Psi_{(1,1,1)}$, as all particles avoid each other due to the vortex factors $\zu_n - \zd_m$. This state has a filling fraction of $\nu=1$ and is a ferromagnet with maximum total $S$ and $S_z = 0$. The total angular momentum $L$ of the state $\Psi_{(m,m,n)}$ can be obtained by counting the order of the corresponding polynomial yielding $L = m N(N-1) + n N^2$.
Excitations away from $\nu = 1$ are quasi-holes and quasi-electrons in solid state systems. In the rotating quantum gas, a quasi-hole corresponds to the addition of a flux quantum, i.e. angular momentum, to the system. In order to construct a quasi-hole spin-singlet wavefunction, Yoshioka \cite{Yoshioka1998} proposed to add a factor introducing $N$ additional flux quanta to $\Psi_{(1,1,1)}$ in order to satisfy the Fock condition \cite{Haldane1988}. The modified wavefunction can be written as
\begin{equation}
\label{eq:skyr}
\Psi_{q}^{(r)} = \Psi_{(q,q,q)} \textrm{per}|M^{(r)}|
\end{equation}
where $\textrm{per}|M^{(r)}|$ is the permanent of a $N \times N$ matrix with elements $M_{i,j}^{(r)} = (\zu_i - \zd_j)^{r}$. 
These wavefunctions describe magnetic skyrmions in the spin degree of freedom. As $rN$ additional quanta of angular momentum are introduced, these states occur at $L = q N(N-1) + q N^2 + rN$. For large $N$, this corresponds to a small correction of the filling fraction away from its original value of $1/q$, and in the many-body context skyrmions can be understood as excitations very close in filling fraction to the $\Psi_{(q,q,q)}$ host state.

\section{Numerical Results}
\begin{figure}[ht]
	\centering
	\includegraphics{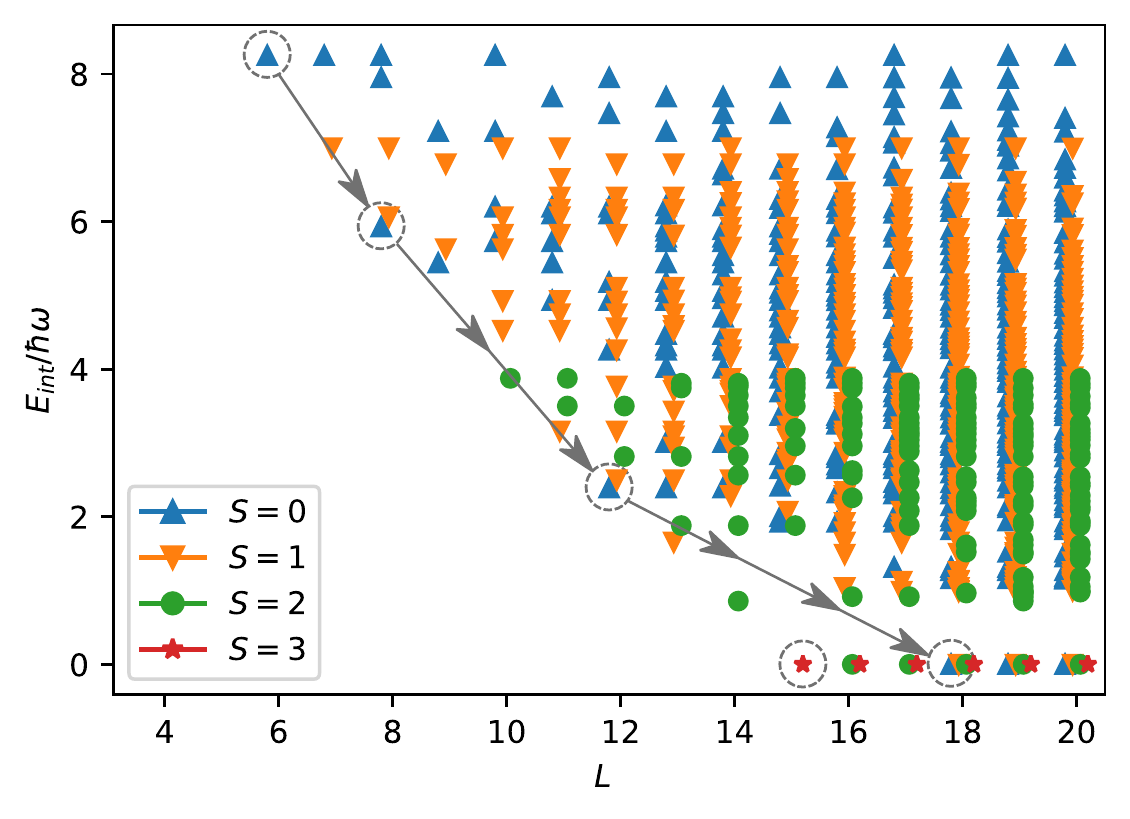}
	\caption{\label{fig:yrast} Yrast spectrum for $N_\uparrow = N_\downarrow = 3$ particles. The eigenstates of $\mathcal{H}_I$ separate into sectors of different total spin $S$. The most strongly correlated states are those with zero interaction egenry $E_\textrm{int}=0$. This requires a minimum angular momentum of $L=15$, for which the total spin is maximal $S=3$. At larger angular momenta, zero interaction energy can also be reached for lower $S$, implying non-trivial spin structures. (Points are offset horizontally for clarity. Circles indicate the states discussed in the text and arrows depict the states traversed in an adiabatic pathway.)}
\end{figure}

We compute the low-energy Yrast spectrum of the interaction Hamiltonian $\mathcal{H}_{I}$, which is shown in \Cref{fig:yrast} for $N_\uparrow = N_\downarrow = 3$ particles. As the total spin $S$ commutes with the Hamiltonian $[\mathcal{H},\vec{S}^2] = 0$, we obtain multiple sectors of total $S$. We focus on the case $N_\uparrow = N_\downarrow$ where all states have total $S_z = 0$.

In order to compare the numerical states obtained in exact diagonalization to analytical trial wavefunctions of the form given in \Cref{eq:Halperin}, we expand the latter polynomials into a second quantized form and compute the overlaps $\mathcal{O} = |\langle \Psi_{trial}|\Psi_{\textrm{ed}}(S,L)\rangle|^2$ numerically. For all states considered in the following we found unity overlap $\mathcal{O} =1$ up to numerical precision for $N_\uparrow = N_\downarrow \leq 4$ particles. In systems with Coulomb interactions, the higher-order Haldane pseudopotentials often lead to only an approximate correspondence between analytic trial states and exact numerical results. Here, for the case of contact interactions, we find that the analytic trial states actually correspond to exact eigentstates.
We first note that all states in the $S=3$ sector suppress interaction energy completely. The first state at $L=15$ is exactly the $\Psi_{(1,1,1)}$ state with filling fraction $\nu=1$. The ground state of the few-fermion system has maximal total spin consistent with quantum Hall ferromagnetism \cite{Yang1994}. 

An even more interesting sequence of states occurs in the $S=0$ sector: The first state at $L=6$ is the trivial $\Psi_{(1,1,0)}$ state described earlier. When angular momentum is transferred into the system, the lower sequence of states (Yrast line) in \Cref{fig:yrast} is traversed within $S=0$.
Moving towards higher $L$, center-of-mass vortices of the form $\Psi_{com}^q = \Psi_{(1,1,0)}  \left(\Zu - \Zd \right)^q $ with c.o.m. coordinates $\Zu = \sum_i z_i / N,\Zd = \sum_i \tilde{z}_i / N$ can be identified at $L=8$ with $q=2$. 

The first state with vanishing interaction energy in this sequence is located at $L=18$. At the cost of larger angular momentum, this state may represent a non-trivial spin structure on top of the ferromagnetic ground state at $|S,L\rangle = |3,15\rangle$. 
Indeed, we find that this is the skyrmion state $\Psi_{1}^{(1)}$, an eigenstate with zero interaction energy. The wavefunction \Cref{eq:skyr}, which was originally conceived for systems with Coulomb interactions on a sphere, accurately describes particles with contact interactions on the disk and we identify the $S=0$ ground states as skyrmions \cite{Yoshioka1989}.

\begin{figure}[t]
	\centering
	\includegraphics[width=\textwidth]{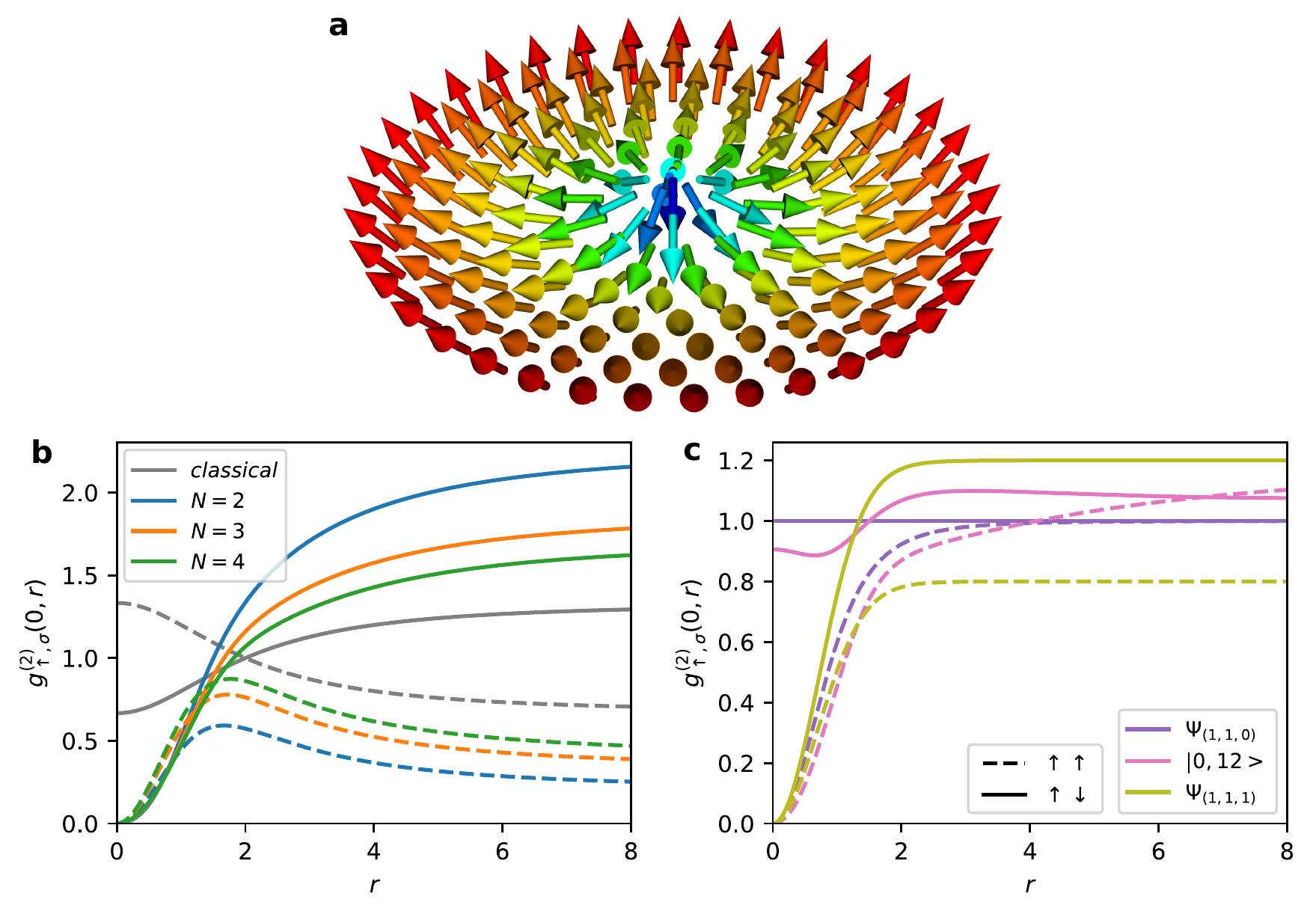}
	\caption{\label{fig:correlations} Skyrmionic spin correlations. (a) Magnetization texture of the classical skyrmion defined in \Cref{eq:classicalskyr}, displaying local ferromagnetic correlations and global spin reversals. (b) Spin-resolved correlation functions for the numerically obtained singlet ground states for different particle numbers.Opposite (equal) spin correlations are shown as solid (dashed) lines. Shown in gray are the semiclassical correlation functions given in \Cref{eq:classicalskyr}. Here the scale of the skyrmion has been set to $\lambda =1$. The correlations of the few-body quantum state are in good qualitative agreement with the semiclassical expectation. (c) Correlation functions for several of the low-energy, low-angular-momentum states in the Yrast spectrum for $N_\uparrow = N_\downarrow = 3$ particles. These states do not exhibit correlations of the type expected for skyrmions.}
\end{figure}

The central feature of skyrmion states are long-range spin textures that can be classified by their topology. Such textures typically occur as excitations on top of ferromagnetic states, where they globally accommodate reversed spins, but locally retain ferromagnetic correlations. Skyrmionic excitations of quantum Hall ferromagnets near $\nu =1$ have been investigated theoretically through construction schemes of the exact quantum mechanical wave function \cite{MacDonald1996}, by semiclassical models \cite{Sondhi1993}, or using trial wavefunctions that interpolate between these two cases \cite{Moon1995}. 

In a semiclassical approximation, the spin texture is represented by a magnetization vector whose spatial variations are slow compared to the microscopic length scale of the system. Skyrmionic spin textures then emerge as the low-energy configurations of the classical field. 

The simplest skyrmion of spatial scale $\lambda$ can be described by a magnetization vector \cite{Sondhi1993} $\mathbf{n} = (n_z, n_r, n_\varphi) = (\cos(\vartheta_r), \sin(\vartheta_r),0)$ with 
\begin{equation}
\label{eq:classicalskyr}
   \cos(\vartheta_r) =  \frac{r^2 - 4 \lambda^2}{r^2 + 4 \lambda^2}.
\end{equation}
\Cref{fig:correlations} (a) shows the magnetization texture of this skyrmion. 
To probe the skyrmionic nature of the $S=0$ ground states we find numerically, we compute their second-order spin correlation functions
\begin{equation}
g^{(2)}_{\sigma, \sigma^\prime} (r,r^\prime) =  \frac{ \left \langle \Psi^\dagger_\sigma(r) \Psi^\dagger_{\sigma^\prime}(r^\prime) \Psi_\sigma(r) \Psi_{\sigma^\prime}(r^\prime) \right \rangle}{ \left  \langle \Psi^\dagger_\sigma(r) \Psi_\sigma (r)\right \rangle \left \langle\Psi^\dagger_{\sigma^\prime}(r^\prime)  \Psi_{\sigma^\prime}(r^\prime) \right \rangle}
\end{equation}
defined in terms of the field operators $\Psi_\sigma(z) = \sum_{m} \phi_m(z) c_{m,\sigma}$ with wavefunctions $\phi_m(z) = z^m e^{-|z|^2/4}/ \sqrt{\pi m!}$. We expect these quantum mechanical correlation functions to reflect the features of the classical spin texture. There are, however, ambiguities associated with a direct comparison, because the classical skyrmion in  \Cref{eq:classicalskyr} does not correspond to a total spin singlet. 

The most meaningful correspondence is achieved by considering an $SU(2)$-invariant ensemble of classical skyrmions obtained by including all global spin rotations of the classical skyrmion in \Cref{eq:classicalskyr} with equal weights for all directions on the unit sphere. The probability for detecting an up (down) spin at a given location $\vec{r}$ in the classical skyrmion is $p_\uparrow=[1+n_z(\vec{r};\theta,\phi)]/2$ ($p_\downarrow=[1-n_z(\vec{r};\theta,\phi)]/2$), where $n_z(\vec{r};\theta,\phi)$ denotes the $z$-component of the classical spin in a skyrmion state rotated by the two spherical angles $\theta,\phi$. Integration over all spin angles $\theta,\phi$ on the unit sphere yields the two-point correlation function of the $SU(2)$ invariant classical ensemble,
\begin{equation}
\tilde{g}_{\uparrow \uparrow}(\vec{r}_1,\vec{r}_2) = \frac{1}{4 \pi} \int_0^{2 \pi} \mathrm{d}\phi \int_0^\pi \mathrm{d}\theta ~ \sin(\theta) \frac{1}{4} [1+n_z(\vec{r}_1;\theta,\phi)] [1+n_z(\vec{r}_2;\theta,\phi)].
\end{equation}
A similar result holds for $g_{\uparrow \downarrow}$, but with a minus sign in the second square bracket. Using $n_z(\vec{r};\theta,\phi) = \vec{e}_r(\theta,\phi+\pi) \cdot \vec{n}(\vec{r};0,0)$ these integrals can be easily evaluated for the classical skyrmion; setting $\vec{r}_1=0$ and $r=|\vec{r}_2|$ we obtain:
\begin{equation}
\label{eq:gclassical}
    \tilde{g}_{\uparrow \uparrow}(r) = \left( 1 + \frac{1}{3} \cos(\vartheta_r) \right), \qquad \tilde{g}_{\uparrow \downarrow}(r) = \left( 1 - \frac{1}{3} \cos(\vartheta_r) \right).
\end{equation}
To compare the semiclassical expectation to a quantum mechanical few-body state, we set its spatial scale $\lambda$ to give zero net magnetization over the area of the corresponding few-body state. We find $\lambda \approx 1$ for $N = 3$ and expect a scaling $\lambda \propto \sqrt{N}$ for larger particle numbers. 

\Cref{fig:correlations} (b) shows the correlations we find for the $S=0$ skyrmion ground state for $N_\uparrow = N_\downarrow = 2$, $3$ and $4$ as well as the semiclassical expectation $\tilde{g}$. Beyond  the correlation hole near the origin, the states show correlations of equal spins on short distances and of opposite spins on longer distances, agreeing remarkably well with the correlations expected semiclassically. Quantitatively, the correlations depend significantly on the particle number, reflecting substantial finite size effects. It is clear, however, that even for small total particle numbers $\sim 10$ the skyrmionic spin texture is recovered. 

We show the spin correlation functions for several other states in the neighborhood of $|S,L\rangle = |0,18\rangle $ in \Cref{fig:correlations} (c). None of these states exhibit correlations consistent with the semiclassical expectation, indicating that such correlations are a clear signature of the skyrmion state. The calculated spin correlation functions therefore further support the identification of the $S=0$ ground states as skyrmions.

In conclusion, ferromagnetic ground states occur in few-fermion rotating systems and ground states in the $S=0$ sector exhibit spin structure akin to skyrmion wavefunctions found in solid state systems. The qualitative similarity of spin textures in systems of vastly different particle numbers is remarkable and highlights the relevance of the few-body approach. We speculate that interesting excitations, such as spin waves on top of the ferromagnet (with $S$ close to maximal) and angular excitations of the skyrmion (at $S=0$ and large angular momentum) can also be identified in the few-body system.

\section{Adiabatic Preparation}
\begin{figure}[ht]
	\centering
	\includegraphics{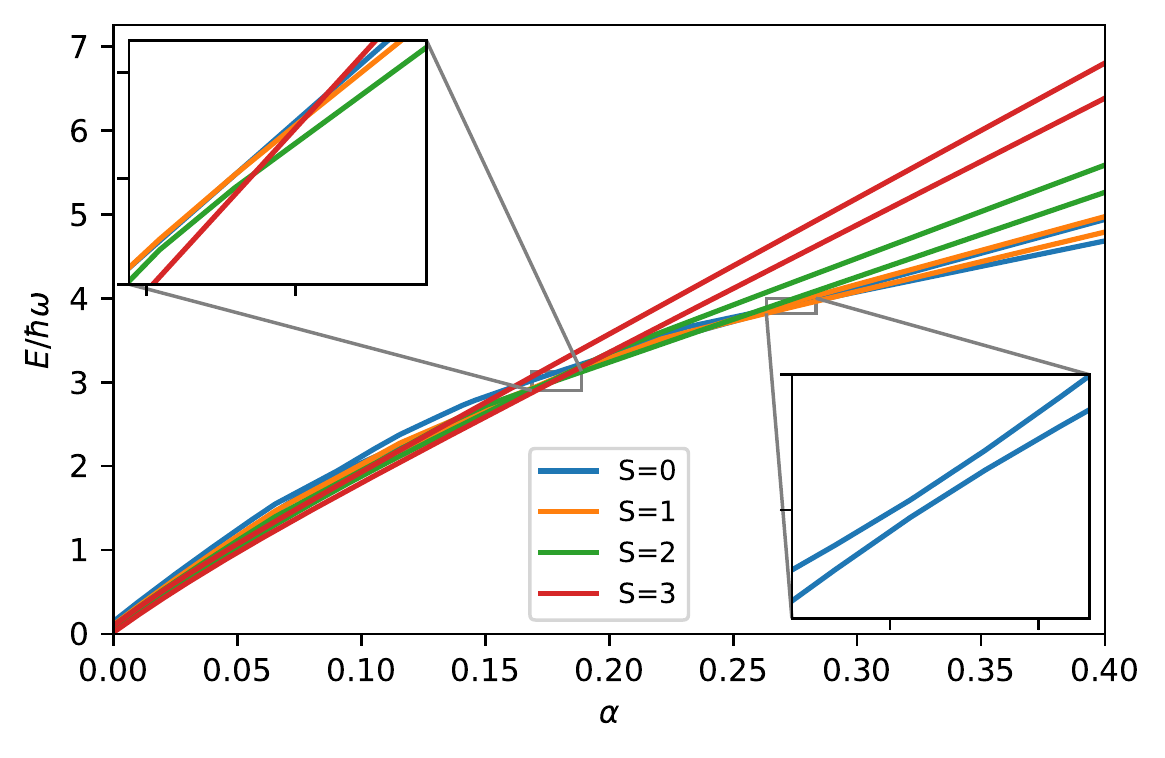}
	\caption{\label{fig:adiabatic} Low energy many-body spectrum for $N_\uparrow = N_\downarrow = 3$ particles for increasing rate of rotation $\alpha \rightarrow 0$ at a trap anisotropy of $\epsilon=0.02$ and interaction strength $\eta=0.25$. Avoided crossings occur within each spin manifold (shown for $S=0$ in the lower inset), while sectors of different $S$ exhibit true crossings (upper inset). The total S is therefore conserved during the adiabatic preparation.}
\end{figure}

We now discuss adiabatic pathways for the preparation of skyrmion-like states. Since the global spin is conserved throughout the state preparation considered here, only a subset of correlated ground states connects to a particular initial state in the resting trap. This is illustrated in \Cref{fig:adiabatic}. Because of its practical relevance, we choose as the starting point of our investigation the state $\Psi_{(1,1,0)}$ with $|S,L\rangle = |0,6\rangle $. From the resting trap, this state can be reached by transferring all fermions to the LLL using the procedure outlined in section \ref{ssec:single}

We asses the feasibility of an adiabatic pathway to the correlated states by calculating the size of the many-body gap $\Delta = E_1 - E_0$ in terms of the ground- and first excited-state energies $E_0, E_1$ within the $S=0$ sector as a function of the rotational frequency $\Omega$.
Working in a rotating frame, we can write \Cref{eq:pert-general} independent of time as
\begin{equation}
\label{eq:pert-rot-general}
\mathcal{H}_p = \sum_{l=1}^{l^*} \epsilon_l \sum_{m,\sigma} v_{m,l} \left( c_{m+l}^{\sigma\dagger} c_{m}^{\sigma} + h.c.\right)
\end{equation}
with $v_{m,l} = 2^{-l/2} (m+l)!/\sqrt{m! (m+l)!}$.

We chose an elliptic perturbation of order $l=2$ tunable strength $\epsilon = \epsilon_2/2$ (see \Cref{eq:pert-rot-general}), such that states within the LLL with a relative angular momentum $\Delta m = 2$ couple as
\begin{equation}
\mathcal{H}_p(\epsilon) = \epsilon \sum_{m,\sigma} \sqrt{(m+2)(m+1)} \left( c^{\sigma\dagger}_{m+2} c_{m}^{\sigma} + h.c. \right).
\end{equation}

\begin{figure}[ht]
	\centering
	\includegraphics{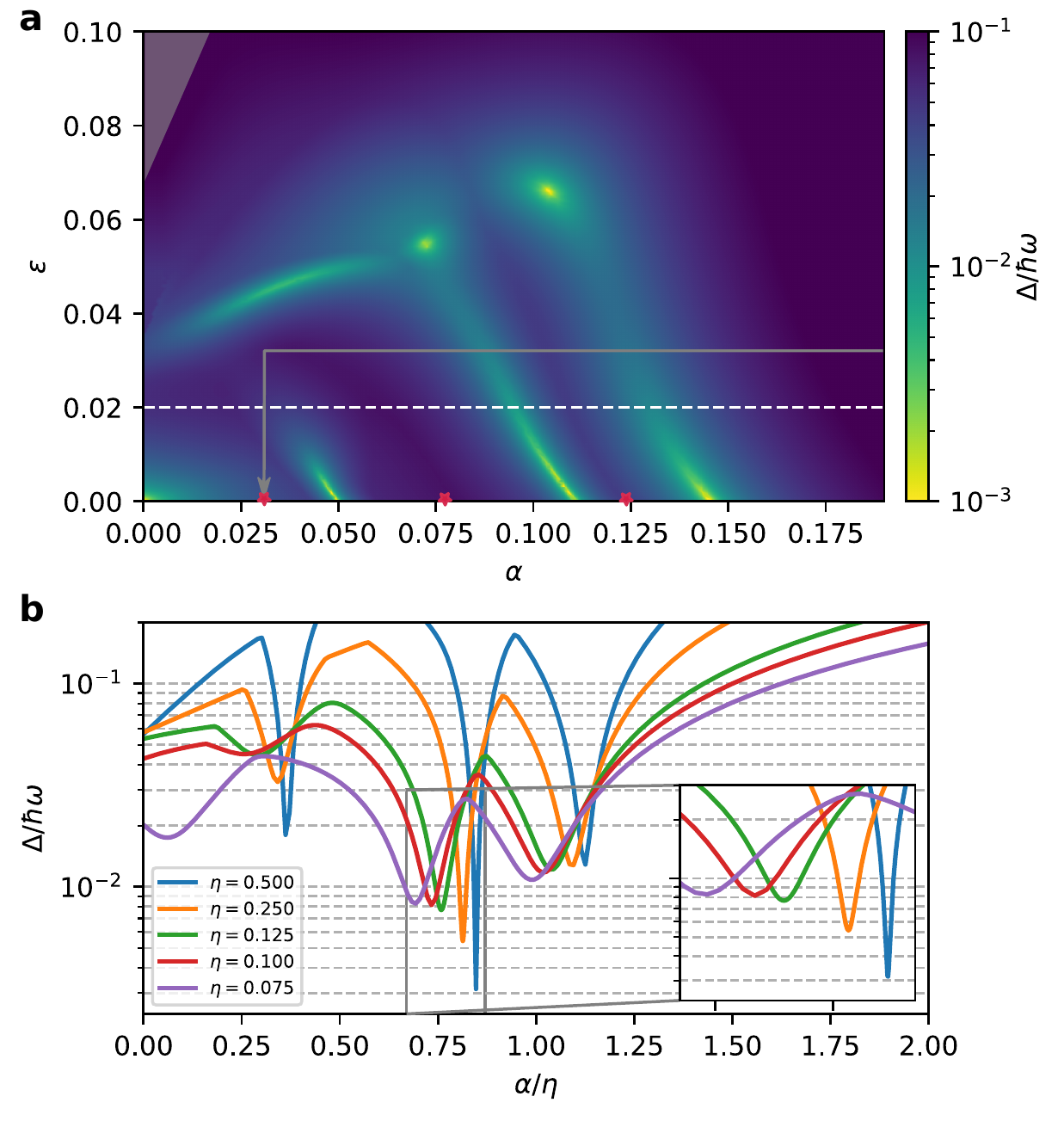}
	\caption{\label{fig:adiabatic2} (a) Size of the many-body gap $\Delta(\alpha,\epsilon)$ in the $S=0$ sector as a function of scaled rotational frequency $\alpha$ and elliptic perturbation strength $\epsilon$ for an interaction strength of $\eta = 0.125$. The sequence of ground states accessible by adiabatic preparation is shown as red stars, while a white dashed line indicates the perturbation strength in (b) and \Cref{fig:adiabatic}. The adiabatic pathway used to estimate the minimal time is shown as a grey arrow. Data in the greyed out area is inaccessible due to numerical instabilities. (b) Scaling of the gap size in the $S=0$ sector for different interactions strengths $\eta$. The gap positions coincide to first order when the rate of rotation is scaled as $\alpha / \eta$.}
\end{figure}

In \Cref{fig:adiabatic2} (a), the resulting gap size in the $S=0$ sector at an interaction strength of $\eta = 0.125$ is shown as a function of these two parameters in a landscape $\Delta(\alpha,\epsilon)$.  For this intermediate interaction strength, the first level crossing occurs at $\alpha \approx 0.15$ and the last at $\alpha \approx 0.05$, i.e. 85\% and 95\% of the trap frequency. To examine the dependence of the gap size and position on the interaction strength more closely, \Cref{fig:adiabatic2} (b) shows a cut of (a) along a line of $\epsilon = 0.02$ for different $\eta$. Because the crossing positions scale with $\eta$ to first order, the x-axis is scaled as $\alpha/\eta$ to approximately make the curves collapse. This scaling highlights the key advantage of working in systems with widely tunable interactions: For large $\eta$, the most interesting states occur already at moderate rotation frequencies, away from the centrifugal limit at $\alpha = 0$. On the other hand, as seen from \Cref{fig:adiabatic2} (b), interaction strengths beyond $\eta \sim 0.1$ significantly reduce the size of the encountered spectral gaps. For an optimal preparation pathway, the interactions strength can be chosen to balance the requirements of final rotation rate and gap size during preparation.

\section{Experimental Implementation \& Detection}

A number of experiments have explored the physics of the lowest Landau level with bulk ultracold gases in rapid rotation \cite{Fetter2009, Fletcher2019}. However, reaching the limit of small filling fraction is difficult in large systems. A promising alternative is to use few-body systems as pioneered in \cite{Gemelke2010,Popp2004}, where high-fidelity preparation and readout schemes are available. 
We now consider a concrete experimental realisation using ultracold $^6$Li with possible experimental parameters shown in \Cref{tab:params}. Our proposed protocol proceeds as follows:
(i) First, a static harmonic trap is loaded with a well-defined number of atoms (per spin component) using the technique established in \cite{Serwane2011}. By choosing an appropriate initial state, the sector of total $S$ can be selected as this is conserved during the whole protocol. The most natural initial state is a non-interacting two-component Fermi sea, which fixes $S=0$. The trap frequencies $(\omega, \omega, \omega_z)$ in the x, y and z direction are chosen to obtain a strong confinement in the $z$ direction and a quasi-2D system with an aspect ratio of $R=\omega_z/\omega > 30$.
(ii) Next, the atoms are transferred into the LLL as described above by adiabatically turning on a rotating trap perturbation. Interactions at this point are still switched off via the Feshbach resonance. (iii) After all particles are in the LLL at an intermediate rate of rotation $\alpha$, interactions are adiabatically increased up to the desired scattering length. The broad Feshbach resonance in $^6$Li allows for very strong interactions of several $1000\,a_0$ corresponding to $\eta \sim 1$, although we have found $\eta \sim 0.1$ to be optimal for the parameters in \Cref{tab:params}. Subsequently, the rotational frequency is increased towards $\alpha \rightarrow 0$ while simultaneously tuning $\epsilon$ according to an optimized pathway through the landscape shown in \Cref{fig:adiabatic2} (a), until the state of desired angular momentum $L$ is reached. 
We estimate the minimal total time of this path following the approach of \cite{Roncaglia2011} in computing the adiabatic condition $\left| \left\langle \psi_0 \left| \partial \mathcal{H}_x / \partial t \right| \psi_i \right\rangle \right| \ll \omega \Delta_i^2$
with the gap $\Delta_i = E_i - E_0$ between the ground- $\psi_0$ and an excited-state $\psi_i$ for both the kinetic Hamiltonian $\mathcal{H}_{\alpha} = \mathcal{H}_0(\alpha)$ and the elliptic perturbation $\mathcal{H}_{\epsilon} = \mathcal{H}_p(\epsilon)$. 
The total minimum preparation time $T$ is then given by a condition $\int F_{\alpha} d\alpha + F_{\epsilon}d\epsilon = \omega T$  on the line integral along a preparation path. Here, $F_x = \Delta^{-2} \left| \left\langle \psi_0 \left| \partial \mathcal{H}_x / \partial x \right| \psi_i \right\rangle \right|$ with $x \in (\alpha, \epsilon)$ is the local adiabatic speed limit. 
We numerically evaluate the line integral for the path shown in \Cref{fig:adiabatic2}, considering the most stringent speed limit arising from the first three excited states in the $S=0$ sector at any point. This yields a minimum time of $T=185 \omega^{-1}$. For the parameters in \Cref{tab:params},this amounts to $T=\SI{30}{ms}$, which is compatible with current experiments. Optimizing the adiabatic pathway or more sophisticated control techniques \cite{Baur2008} may reduce this time even further.
To detect the previously prepared states, a time-of-flight expansion also termed the ``wavefunction microscope" \cite{Read2003} can be employed. In this scheme, the wavefunction is magnified while retaining all correlations and it can be imaged through a free-space single-particle imaging technique developed in \cite{Bergschneider2018}. Such imaging methods can detect the positions and spins of individual particles and enable the experimental measurement of spin-resolved second-order correlation functions shown in \Cref{fig:correlations}.

Information beyond the correlation functions is contained in the individual quantum projective measurements. This allows, for example, for an unbiased comparison with theoretical predictions using machine-learning \cite{Bohrdt2019, Rem2019}. For skyrmions, a particularly interesting possibility is to apply position-dependent spin-rotations before the measurements, allowing to study the spatial structure of the spin textures. Furthermore, by measuring density fluctuations over a finite region, the incompressibility of the state can be directly demonstrated. To confirm that the correct total spin sector $S=0$ of the skyrmion wavefunction has been prepared, one can measure the total $S^z$ component after a random rotation of the measurement basis for the spins: if for all such rotations $S^z_{\rm tot}=0$, the prepared state is a singlet.

\section{Conclusion}
In summary, we show that a small sample of ultracold fermionic atoms can host stable magnetic excitations known as skyrmions.
\begin{table}[ht]
	\centering
	\begin{tabular}{lcr}
		\br
		Parameter & Symbol & Value\\
		\mr
		axial trap frequency & $\omega_z$ & $2\pi \times \SI{30}{kHz}$\\
		z-confinement h.o. length & $l_z$ & \SI{240}{nm}\\ 
		magnetic length & $l$ & \SI{1.3}{\mu m}\\ 
		scattering length & $a_s$ & \SI{100}{a_B} \\
		interaction parameter & $\eta$ & $0.1$\\
		radial trap frequency& $\omega$ & $2\pi \times \SI{1}{kHz}$\\
		aspect ratio & $R$ & $30$\\
		perturbation strength & $\epsilon$ & $0.02$\\
		gap size & $\Delta$ & $2\pi \times \SI{10}{Hz}$\\
		\br
	\end{tabular}
	
	\caption{\label{tab:params}Experimental parameters. Typical experimental parameters that may be realized in a cold-atom setup using $^6\textrm{Li}$ as in \protect \cite{Serwane2011, Bergschneider2018}.}
\end{table}
The flat bands required for QH physics in the LLL can be reached through rapid rotation of an optical trap with precise control over the shape and strength of an engineered perturbation. The size of the many-body gap allows for the preparation of highly correlated QH states for realistic parameters.
Our results are also relevant for other synthetic quantum systems where multi-component quantum Hall systems may be realizable, for example ultracold atoms in optical lattices \cite{Sorensen2005,He2017} or photonic systems with engineered interactions \cite{Clark2019}. In the future, ultracold fermionic atoms with dipolar or other long-range interactions may be used to access systems with richer pseudopotentials \cite{Baranov2002}. This will enable the study of the interplay between spin textures and fractional quantum Hall physics reminiscent of the Coulomb interaction encountered in solid state.

\ack
We thank Eugene Demler, Selim Jochim, and Matteo Rizzi for insightful discussions and Eric Tai for help with the numerical calculations. This work has been supported by the ERC consolidator grant 725636, the Heidelberg Center for Quantum Dynamics, the DFG Collaborative Research Centre SFB 1225 (ISOQUANT) and by the Deutsche Forschungsgemeinschaft (DFG, German Research Foundation) under Germany's Excellence Strategy -- EXC-2111 -- 390814868. PMP acknowledges funding from the from the Daimler and Benz Foundation.

\newpage
\section*{References}
\bibliographystyle{my-iopart-num}
\bibliography{skyrmion}

\end{document}